\documentstyle[multicol,aps,psfig]{revtex}
\begin{document}

\draft

\title{A coarse grained model for granular compaction and relaxation}

\author{D. A. Head\cite{em1} and G. J. Rodgers\cite{em2}}

\address{Institute of Physical and Environmental Sciences,
Brunel University, Uxbridge, Middlesex, UB8~3PH, United Kingdom}

\date{\today}

\maketitle

\begin{abstract}
We introduce a theoretical model for the compaction of granular
materials by discrete vibrations which is expected to hold when
the intensity of vibration is low.
The dynamical unit is taken to be clusters of granules that belong
to the same collective structure.
We rigourously construct the model from first principles and
show that numerical solutions compare favourably with a
range of experimental results.
This includes the logarithmic relaxation towards
a statistical steady state, the effect of varying the intensity
of vibration resulting in a so-called ``annealing'' curve,
and the power spectrum of density fluctuations in the
steady state itself.
A mean field version of the model is introduced which
shares many features with the exact model and is open
to quantitative analysis.
\end{abstract}


\begin{multicols}{2}
\narrowtext

\section{Introduction}
\label{sec:intro}

Extrapolating bulk properties from the underlying
microscopic dynamics is generally more difficult with granular materials
than with gases, a difficulty that
has been attributed, at least in part, to the lack of thermal 
averaging~\cite{rev1,rev2}. Unlike molecules, granules are static at room
temperature and so cannot explore phase space without some
external impetus. For example, consider a column of loosely
packed granules in a cylindrical container,
where {\em loosely packed} means that there are typically
large gaps or {\em voids} between neighbouring granules. It is energetically 
favourable for the granules to collectively reorganise to a state
which minimises these voids, since a more compact column will have
a lower centre of gravity and hence a lower
potential energy. That this does not occur spontaneously is a
direct consequence of the lack of thermal motion.
One way to allow the column to evolve
is simply to {\em tap} or otherwise
perturb the container, thus giving the granules a small amount
of kinetic energy with which to rearrange.
This process has been studied empirically in the context of
industrial applications~\cite{industrial}, but only recently
have attempts been made to try to understand
the fundamental dynamics involved.

Mehta~{\em et.al.}~\cite{mehta1,mehta2,mehta3} employed a 
non-sequential Monte Carlo algorithm to simulate the process
on a microscopic level.
{\em Non-sequential} means that granules are
allowed to move and settle simultaneously, which is important in this 
context since it allows for the cooperative reorganisation of 
granule-granule contacts. These simulations predict that granular media
should relax on two time scales, corresponding to individual granule 
motion and collective processes respectively. However, 
this is not in accord with the experimental work of
Knight~{\em et.al.}~\cite{exp}.
They measured the rate of compaction in a column of monodisperse glass 
beads that was subjected to discrete vertical vibrations.
The plot of density against the number of vibrations was found
to be best 
described by \mbox{$\rho(t)\sim(\log t)^{-1}$}, where the time ordinate $t$ 
is proportional to the number of taps. One possible reason for the 
discrepancy between the simulations and the experiments may simply be 
that the regimes of vibration intensity studied were different. The smallest 
vibration considered in the simulations corresponds to a 5{\%} increase in 
volume at every tap, which is much more than the 
experiments involved.

A number of models embracing a variety of theoretical approaches
have been introduced
to try and account for the experimental findings.
Of those we are aware of, one is a phenomenological macroscopic
model~\cite{linz}, but the remainder are all microscopic in
nature. The slow relaxation has been attributed
by Ben-Naim~{\em et.al.} to the large number of reconfigurations
required to bring enough small voids together
to make one void large enough to absorb another
granule~\cite{rsa1,rsa2,exp2}.
de Gennes also chose to focus on the voids and found that
a Poisson distribution of void sizes could give rise to the expected
inverse logarithmic relaxation~\cite{de_Gennes}.
Coglioti~{\em et.al.} have introduced a lattice model in which each 
granule can be in one of two states with each state corresponding to a 
different geometrical orientation~\cite{tetris,frus1}.
The motion between neighbouring granules is 
constrained by their relative orientations, hence the rate of relaxation 
in their model
is governed by a form of {\em geometrical frustration}.

In this paper, we introduce a model for granular compaction which is 
neither macroscopic nor microscopic but instead lies somewhere 
between these two extremes.
It is {\em coarse grained} in that it takes {\em clusters} of 
granules as its dynamical unit rather than individual granules.
This approach is based on the picture of granular 
interactions described by Mehta~{\em et.al.} in 
relation to their simulations~\cite{mehta1,mehta2,mehta3},
except that here we are interested in the limit of
weak vibrations.
The resulting model is strikingly similar to one 
already devised by Bak and Sneppen in a wildly different context, that of 
biological evolution~\cite{BS,BSRev}.
In Sec.~\ref{sec:model} the model is described in detail and its physical
basis is explained. Careful consideration is given to the
range of validity of our assumptions.
Results of numerical simulations are compared to the experimental
findings in Sec.~\ref{sec:numer}.
The exact solution of a mean field version of the model is investigated
in Sec.~\ref{sec:mft}.
Finally, with give a summary of the model in
Sec.~\ref{sec:summ}.

\section{The Model}
\label{sec:model}

\begin{figure}
\centerline{\psfig{file=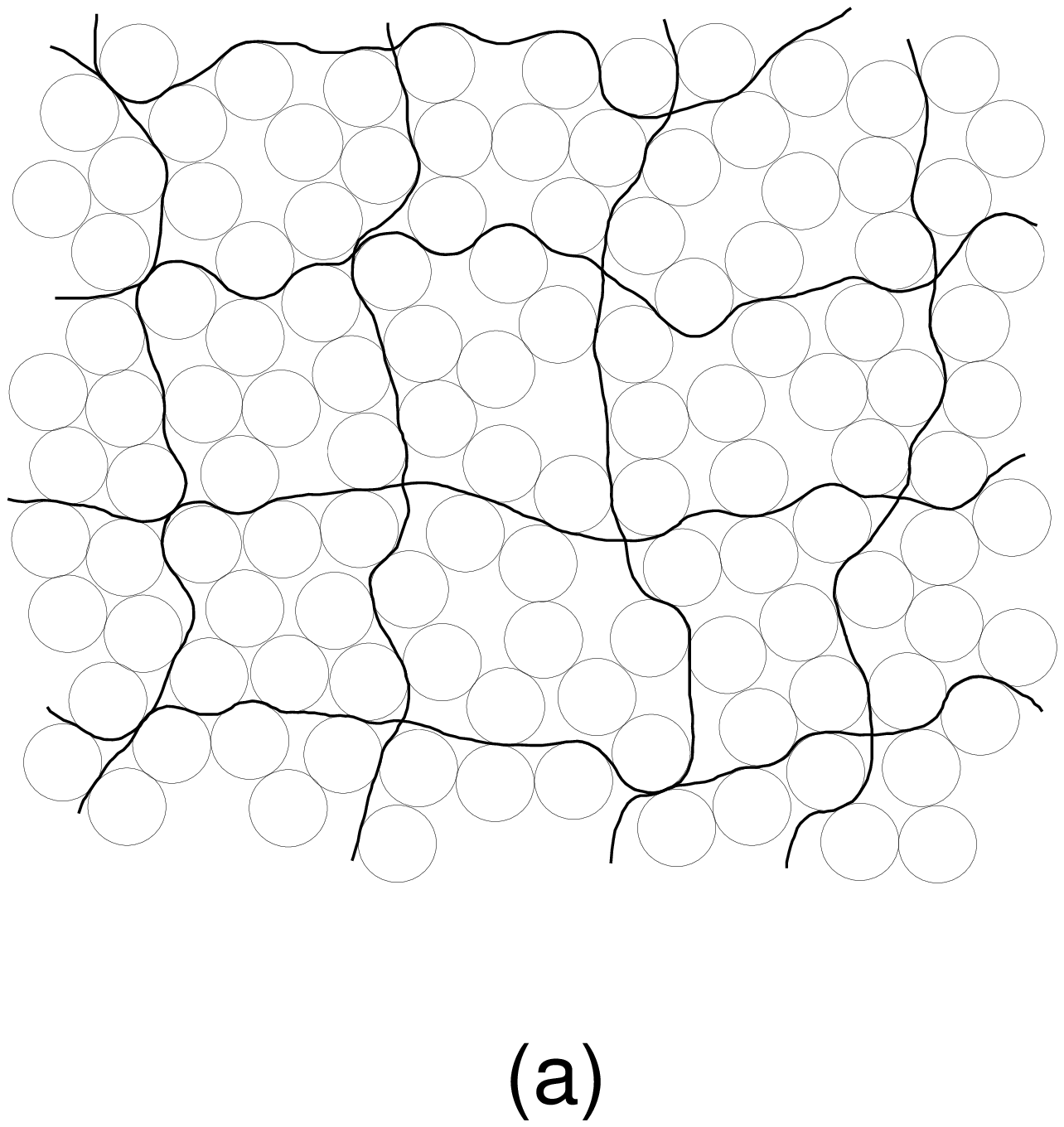,width=2in}}
\centerline{\psfig{file=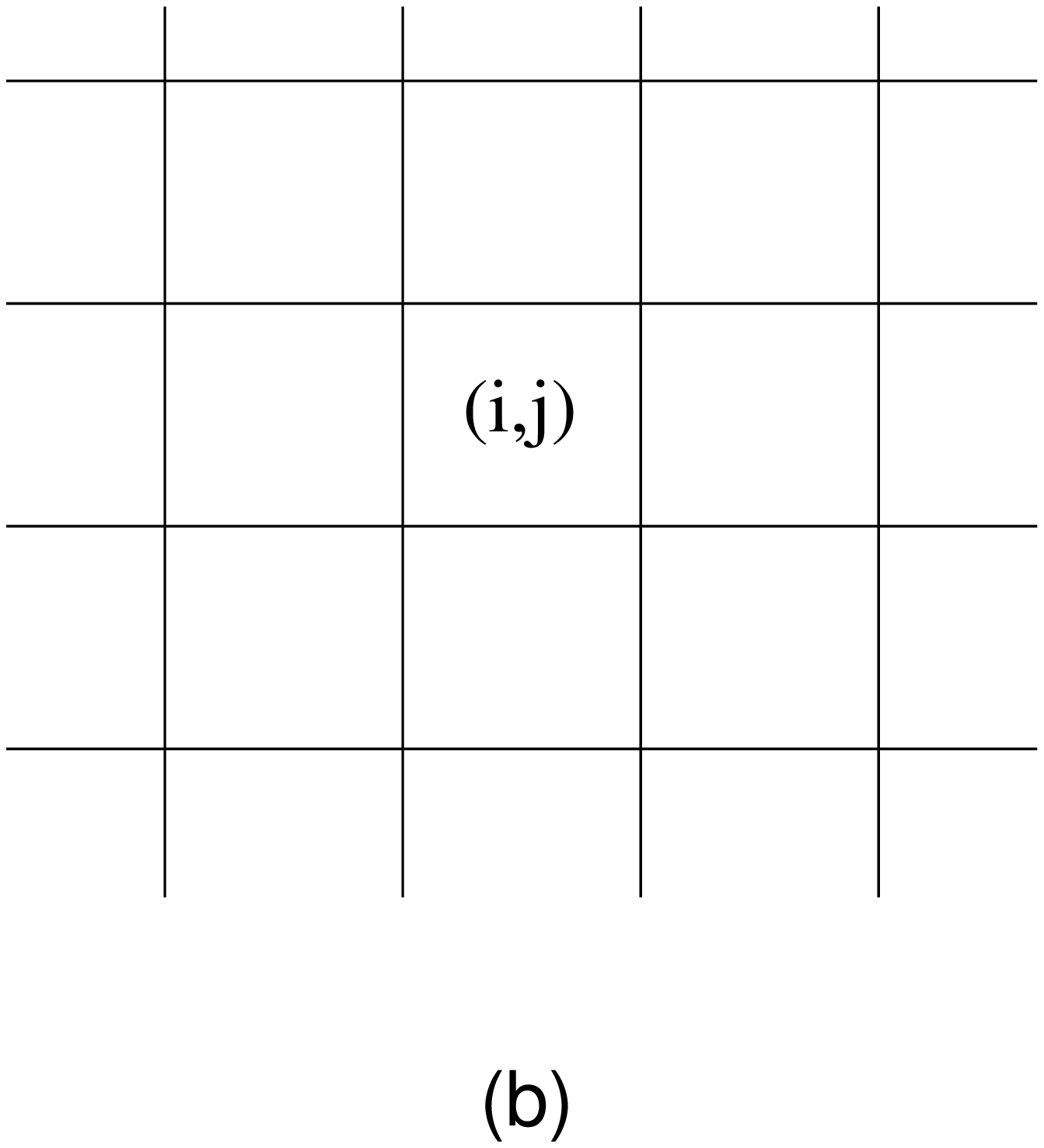,width=2in}}
\caption{An example of the process of subdividing granular media into
local clusters, given here for the case of two dimensions.
(a) A collection of circular granules separated into clusters.
The thick lines represent boundaries between neighbouring clusters.
(b) The corresponding lattice representation. Each site
\mbox{$(i,j)$} denotes a single cluster.}
\label{f:clusters}
\end{figure}

Mehta~{\em et.al.}
picture the granular media as being subdivided into local
clusters, as in Fig.~\ref{f:clusters}(a),
where a cluster is defined as a group of granules belonging
to the same multi-particle potential well~\cite{mehta1,mehta2,mehta3}.
A vibration with an intensity equivalent to 
the binding energy of a granule to its well causes that granule to be 
ejected and move independently of the others.
Under weaker vibrations, all the granules remain in the well but
still reorganise collectively, albeit on a slower time scale to
individual particle motion.
Although this description seems to be valid for the range of 
intensities of vibration considered in their simulations, it clearly
fails for the much lower intensities relevant to the
experiments~\cite{exp,exp2}.
We believe that the picture is essentially correct but needs to be 
modified to describe the behaviour of the system deep in the collective 
relaxation regime.
To do this, we first need to closely analyse exactly what is meant
by a multi-particle potential well.

\begin{figure}
\centerline{\psfig{file=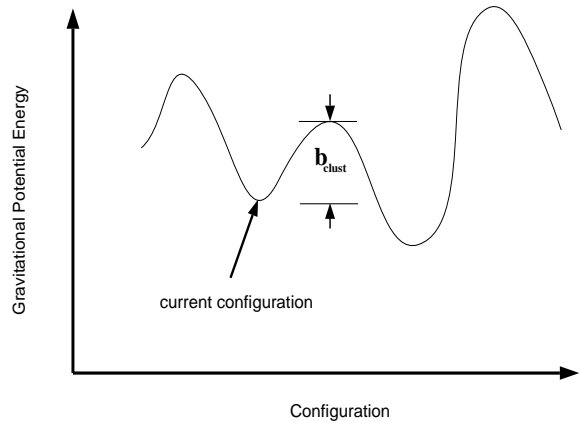,width=3in}}
\caption{Schematic example of a potential energy landscape for an
ensemble of granules in configuration space. The ensemble currently lies
at the local minimum marked. The smallest barrier to an adjacent minimum
has a height of~$b_{\rm clust}$.}
\label{f:landscape}
\end{figure}

Any given configuration of an ensemble of particles can be represented
by a single point in the space of all possible configurations.
Each allowed configuration has a well defined potential
energy, and so the time evolution of the ensemble under gravity
can be described by a walk in configuration space over a
potential energy {\em landscape}.
Now, the preferred state for each individual granule is simply
resting at the bottom of the container.
If the granules did not interact, then the ensemble would
trivially evolve to the global minimum with every
granule in its preferred state, i.e. all resting on the bottom.
Of course, real granules do interact, and one granule moving
downwards will inevitably push some of the surrounding granules
upwards slightly.
The ensemble is thus {\em frustrated} in that it cannot
simultaneously satisfy each granule's tendency to move downwards.
In terms of the potential energy landscape, this frustration
results in a rugged landscape with many local minima
separated by barriers of various heights.
A schematic example is given in Fig.~\ref{f:landscape}, where
for clarity we have compressed the entire configuration space
onto a single axis.

The ensemble will be at a local minimum between perturbations.
The effect of the perturbation is to move the ensemble
to a point higher up on the landscape before it again relaxes, possibly
to a different minimum.
For the low-energy perturbations
we are concerned with here, the ensemble will usually move between nearby
minima and consequently only a small number of granules will change
their position or orientation.
Following Mehta~{\em et.al.}
we assume that these granules typically belong to some sort
of collective structure, such as an arch or bridge.
Thus the system can be subdivided into localised
{\em clusters}, where a cluster is now defined as
the {\em unit of collective reconfiguration}.
Furthermore, we map the system onto a lattice in which every site
corresponds to a single cluster, as in Fig.~\ref{f:clusters}.
This lattice representation is implicitly static and so will not be valid
if there is any form of global motion in the system, such as convection
or surface flow, although it should still hold if there is only a limited
amount of local motion. Large perturbations will involve
reorganisation on a system-wide scale and the rapid
rearrangement of cluster boundaries,
so the lattice representation is again expected to fail
in such situations.

We now have a lattice of clusters, each of which move on their
own individual potential energy landscapes.
During the perturbation, each cluster is kicked to a
point higher up on its landscape,
and those that subsequently relax to a new minimum
have collectively reconfigured.
When a cluster reconfigures the contacts between it and adjacent
clusters will be redistributed in a highly non-trivial manner,
the pattern of stress lines will be locally distorted and the
boundaries between adjacent clusters may shift slightly to
accommodate different granules.
As a consequence, there will be a significant change in the landscapes
of the cluster itself and those near to it.
In particular, we note that the heights of barriers between
minima will change.
It may seem possible for one of the nearby clusters
to move a significant distance on its new landscape before finding
a minimum, effectively constituting another reconfiguration event.
However, this contradicts the definition of a cluster as the fundamental
unit of collective reconfiguration, since
any two clusters that interact in this way should have been treated as
a single cluster in the first place.
Thus it can safely be assumed that nearby clusters will not reconfigure,
although the heights of barriers in their landscapes will still change.

Significant progress can be made if we do away with the landscapes altogether
and just deal with the heights of barriers between minima instead.
Indeed, as we are only interested in the limit of weak perturbations,
we can go one step further and disregard all but the {\em smallest} barrier,
since this will almost always be the one that is involved anyway.
Each reconfiguration is assumed to alter the landscapes in such a complicated
manner that, to good approximation, the height of a barrier can be taken to be
a random number drawn from a suitable probability distribution.
Although this distribution is in general
unknowable, we have found the model to be robust to a variety of
different choices, including uniform, exponential and Gaussian
({\em robustness} means that the essential behaviour of the system
remains unchanged with respect to the modifications tried).
We subsequently use the uniform probability distribution $P(b)$
for barrier height~$b$, where

\begin{equation}
P(b) = \left\{
          \begin{array}{ll}
             1 & \mbox{for $b\in[0,1]$, and} \\
             0 & \mbox{otherwise.}
          \end{array}
       \right.
\label{probdist}
\end{equation}

Consider now the effect of the external perturbation on just a single cluster
with a barrier height of $b_{\rm clust}$.
Suppose that the effect of the perturbation is for the cluster to gain an
energy of $e_{\Gamma}$ and to move to a corresponding point higher up
on its landscape.
If \mbox{$e_{\Gamma}<b_{\rm clust}$}, the cluster cannot cross
even its lowest barrier
and so we can be sure that it will relax to the same minimum
that it was at before.
However, if \mbox{$e_{\Gamma}\geq b_{\rm clust}$} then there is a non-zero
probability that the cluster will reconfigure.
We take this probability to be of the form

\begin{equation}
\left\{
         \begin{array}{ll}
\propto \rm{exp} \left\{ -\mu \left(
\frac{e_{\Gamma}}{e_{\Gamma}-b_{\rm clust}} \right) \right\}
                 & \mbox{for $e_{\Gamma} > b_{\rm clust}$,} \\
             = 0 & \mbox{for $e_{\Gamma}\leq b_{\rm clust}$,}
         \end{array}
\right.
\label{crossprob}
\end{equation}

\noindent{where $\mu$ is a dimensionless constant.
This may appear to be a somewhat arbitrary choice, but a number of
variations with a suitable cut-off at~\mbox{$e_{\Gamma}=b_{\rm clust}$}
were tried, and no essential difference in system behaviour was observed.
The choice of~(\ref{crossprob}) was made since it is exponential in form,
implying some sort of underlying Poisson process, and it has the correct
asymptotics for \mbox{$e_{\Gamma}\rightarrow b_{\rm clust}$} and
\mbox{$e_{\Gamma}\rightarrow\infty$}.
}

When the container is vibrated, the associated energy impulse is distributed
in some undefined manner to all the clusters in the system.
We have observed little qualitative difference arising from distributing
this energy stochastically and henceforth assume that each
cluster receives the same energy $e_{\Gamma}$.
It should be clear from~(\ref{crossprob}) that the cluster with the
smallest barrier in the system, say of height~$b_{\rm{min}}$, is the most
likely to reconfigure. With this observation, we can make
a further simplification that also makes little difference to
the system behaviour,
which is to assume that the cluster that reconfigures first is {\em always}
the one with the barrier height of $b_{\rm{min}}$.
Thus there is no longer any need to simulate
every perturbation until the cluster reconfigures,
we can instead just reconfigure the cluster immediately and advance
the time by an amount \mbox{$\delta t$}, where

\begin{equation}
\delta t \propto
\rm{exp}\left\{\mu\left(
\frac{e_{\Gamma}}{e_{\Gamma}-b_{\rm{min}}}\right)\right\}\:.
\label{timestep}
\end{equation}

\noindent{This is the expected number
of perturbations of energy $e_{\Gamma}$ required until the cluster with
barrier height~$b_{\rm{min}}$ reconfigures, and is the reciprocal
of~(\ref{crossprob}).
For $b_{\rm min}\leq e_{\Gamma}$, $\delta t$ is taken to be infinite.
}

We are now in a position to describe the model algorithmically.
The granular media is represented by a lattice, each site of which
corresponds to a unit of collective reconfiguration, ie. a cluster.
The model is robust to variations in lattice connectivity, so
without loss of generality we choose a simple cubic array.
Each cluster $(i,j,k)$ has an associated potential energy barrier against
reconfiguration, $b_{ijk}$, drawn from the probability
distribution $P(b)$ given in~(\ref{probdist}).
The external perturbation takes the form of an energy impulse
distributed uniformly throughout the system, each cluster receiving
an amount~$e_{\Gamma}$.
At each algorithm step, the cluster with the smallest barrier in the
system, $b_{\rm{min}}$, is found.
If \mbox{$e_{\Gamma}\leq b_{\rm{min}}$} then the perturbation is too weak to
cause any reconfiguration events, the system is frozen and the simulation
is complete.
If \mbox{$e_{\Gamma}>b_{\rm{min}}$}, the cluster in question and the 6
clusters adjacent to it are reconfigured - that is, their barriers are redrawn
from the same probability distribution as before.
The real time is increased by an amount \mbox{$\delta t$}
defined in~(\ref{timestep}),
and the simulation moves on to the next algorithm step.
Note that we do not employ periodic boundary conditions, instead
clusters at the faces, edges or corners of the lattice simply have
5, 4 or 3 adjacent clusters, respectively.

Numerical solutions of the model are presented in the following section.
For now, we would like to remark upon the strong similarity between this
model and a model of biological evolution already devised by Bak and
Sneppen~\cite{BS}.
The lattice sites in their model represent different {\em species},
each of which is assigned a barrier against {\em mutation} corresponding
to the smallest barrier between local optima on a rugged {\em fitness}
landscape.
The species mutate and interact with adjacent species in much the same way
that clusters reconfigure and interact with adjacent clusters in our model.
The primary difference between the models is that, whereas clusters
cannot move higher than $e_{\Gamma}$ on their potential energy landscapes,
corresponding to the strength of the external impulse, species are subject
to no such energetic constraints (there is no such thing as the
``conservation of fitness'')
and move around their fitness landscapes spontaneously.
As long as this difference is borne in mind, we can draw upon the plethora
of results already accumulated for the evolution model in analysing our
model of compaction (for a review, see~\cite{BSRev}).

\section{Comparison to experiments}
\label{sec:numer}

We begin by describing the numerical solution of the model for
a system comprising of $N$ clusters.
The distribution of barrier heights, $Q(b)$, is defined such that a
proportion $Q(b)\delta b$ of the clusters have a barrier height in the range
$b$ to~$b+\delta b$.
As the system evolves, $Q(b)$ exhibits two qualitatively different regions,
one for large~$b$ and one for small~$b$.
Large barriers have either not been touched since the simulation began,
or (more likely) they have been redrawn from the uniform distribution
$P(b)$ as the consequence of an adjacent cluster reconfiguring.
As such, $Q(b)$ for large $b$ must also be uniform, except for statistical
fluctuations.
The situation is more complicated for small barriers since there is now
the added possibility of being selected as the minimum of the system.
Very small barriers are unlikely to last long and so $Q(b)$ tails off to
zero as $b\rightarrow0$.
The boundary between these two regions is given by the {\em gap} function
$G(t)$, which is the largest barrier height that has ever been
the minimum of the system.
Finding the minimum barrier and giving it a new
value can be viewed as a flux from the region \mbox{$b\leq G(t)$}
to the region \mbox{$b>G(t)$}.
When there are no barriers left in the region \mbox{$b\leq G(t)$},
larger barriers will be
selected as the minimum and so $G(t)$ will increase.
If there were no interactions, there would only be this
unidirectional flux
and $G(t)$ would slowly approach~1 as \mbox{$t\rightarrow\infty$}.
However, with interactions there is also a flux in the reverse direction,
from \mbox{$b>G(t)$} to \mbox{$b\leq G(t)$}, corresponding to the new
values given to the barriers of adjacent clusters.
Hence $G(t)$ in fact approaches a constant value
\mbox{$b^{*}\in(0,1)$}, where $b^{*}$ is a function of the lattice
connectivity and the system size~$N$.

\begin{figure}
\centerline{\psfig{file=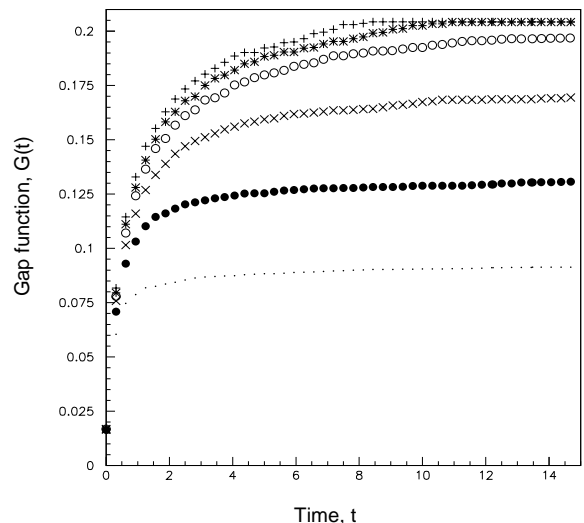,width=3in}}
\caption{Plot of the gap function $G(t)$ for various values of $e_{\Gamma}$,
for a $40\times40\times40$ lattice.
Key:~Plus signs, $e_{\Gamma}=0.4$. Asterixes, $e_{\Gamma}=0.3$.
Open circles, $e_{\Gamma}=0.25$. Crosses, $e_{\Gamma}=0.2$.
Filled circles, $e_{\Gamma}=0.15$. Dots, $e_{\Gamma}=0.1$.
Note that in this and all subsequent plots we have taken the time step to be
\mbox{$\delta t={\rm exp}\{e_{\Gamma}/(e_{\Gamma}-b_{\rm min})\}/N$},
where $N$ is the system size, so the units on the time axis are arbitrary.
}
\label{f:gap}
\end{figure}

We have not yet considered the effect of the parameter~$e_{\Gamma}$.
This appears in the equation for~$\delta t$, the time step between
successive reconfiguration events, which also depends on
the current value of the minimum barrier~(\ref{timestep}).
It can be seen from~(\ref{timestep}) that~$\delta t$
becomes singular when the minimum barrier
is greater than or equal to $e_{\Gamma}$. 
If~\mbox{$e_{\Gamma}>b^{*}$} then this can never happen, since the
minimum fluctuates between $0$ and $G(t)$, and
\mbox{$G(t)\rightarrow b^{*}$} as \mbox{$t\rightarrow\infty$}.
Accordingly the system approaches a statistical steady state in which
$\delta t$ fluctuates around some constant value.
By contrast, if~\mbox{$e_{\Gamma}<b^{*}$} then it now becomes possible
for $G(t)$, and hence also the minimum,
to take values close to $e_{\Gamma}$.
As it does so, $\delta t$ will diverge
and the system will freeze into a state in which
every cluster has a barrier greater than $e_{\Gamma}$ and
there can be no further reconfigurations.
An example of how $G(t)$ depends on $e_{\Gamma}$ is given in Fig.~\ref{f:gap}
for \mbox{a $40\times40\times40$} lattice, for which $b^{*}\approx0.21$.

The model has so far been described in terms of the energy impulse per
cluster $e_{\Gamma}$ and the barrier distribution $Q(b)$.
However, the experimental results were given in terms of
an {\em acceleration} parameter~$\Gamma$ and the {\em density}~$\rho$.
Before comparing the model with the experimental results,
we must first consider how these two
sets of quantities are related.
We start with $e_{\Gamma}$ and~$\Gamma$.
The acceleration parameter~$\Gamma$ is defined as the peak
acceleration during the perturbation scaled by gravity,
\mbox{$\Gamma=a_{\rm max}/g$}.
This was also found to be the relevant parameter for the stability
of a bead heap under vibration~\cite{beadheap}. 
Although it seems reasonable that a higher $\Gamma$ should mean more energy
is distributed throughout the system and hence a higher $e_{\Gamma}$,
the precise relationship is likely to be very complex and we have been
unable to derive a formula relating the two.
Instead we simply assume that, for the small vibrations considered here,
the relationship is approximately linear, \mbox{$e_{\Gamma}\propto\Gamma$}.

Trying to quantify the relationship between the barrier distribution and
density is more problematic since a potential energy barrier is an
intrinsically abstract concept.
Nonetheless, a rough formula can be derived as follows.
Consider an individual cluster with a barrier $b_{\rm clust}$ and density
$\rho_{\rm clust}$.
The cluster's horizontal cross sectional area is assumed to remain
roughly constant throughout the compaction process,
so the typical vertical separation between the granule
centres will be inversely proportional to~$\rho_{\rm clust}$.
The cluster cannot reconfigure unless this vertical separation
is increased to the order of the granule diameter, thus allowing
the granules to move over one another.
Since the granule diameter is constant, the change in height required
for reconfiguration will also depend inversely upon~$\rho_{\rm clust}$.
The potential energy gained by a particle is, of course, proportional
to its height increase, so $b_{\rm clust}$
also varies inversely with $\rho_{\rm clust}$.
Extrapolating this result over the entire system amounts to
finding the mean barrier height~$\bar{b}$, so finally we have

\begin{equation}
\bar{b}\sim\rho^{-1}\:.
\label{ebartorho}
\end{equation}

\noindent{This derivation is simplified in that, for instance,
it does not incorporate the effect of adjacent clusters on the value
of $b_{\rm clust}$.
We expect it to work for overall trends in density but not
for small fluctuations.}

\begin{figure}
\centerline{\psfig{file=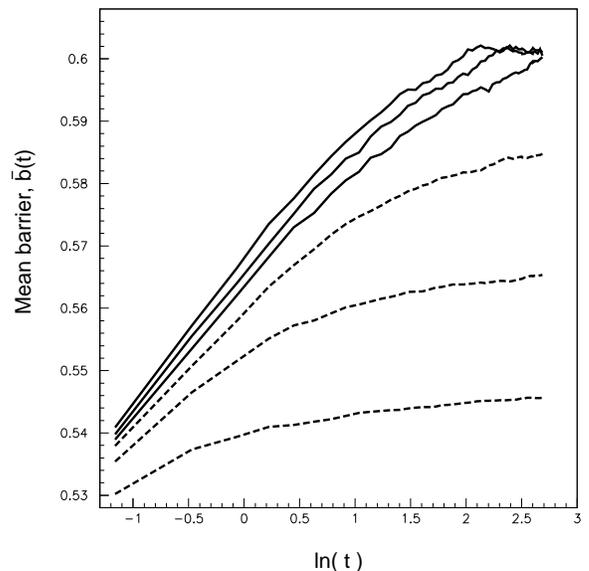,width=3in}}
\caption{$\bar{b}(t)$ vs. $\ln t$ for a range of values of~$e_{\Gamma}$.
The data was taken from single runs on a $40\times40\times40$ lattice,
for which \mbox{$b^{*}\approx0.21$}.
From top to bottom, the values of $e_{\Gamma}$ are:
0.4, 0.3, 0.25, 0.2, 0.15, 0.1.
Solid lines have been used for \mbox{$e_{\Gamma}>b^{*}$} and dashed
lines have been used for \mbox{$e_{\Gamma}<b^{*}$}.
}
\label{f:relax}
\end{figure}

We are now in a position to test the model against the experimental
results.
As mentioned in the introduction, the density was experimentally found
to relax inverse logarithmically with time,
\mbox{$\rho(t)\sim(\log t)^{-1}$}~\cite{exp}.
From~(\ref{ebartorho}) the
corresponding relationship in terms of the mean barrier height is
therefore \mbox{$\bar{b}(t)\sim\log t$}, which will
show up as a straight line on a graph of $\bar{b}(t)$ vs $\log t$.
Such a graph is given in Fig.~\ref{f:relax} for a range of values 
of~$e_{\Gamma}$.
Linear behaviour is apparent over a broad range of densities
for \mbox{$e_{\Gamma}>b^{*}$}, confirming
logarithmic relaxation towards the statistical steady state.
For \mbox{$e_{\Gamma}<b^{*}$}, the relaxation is initially logarithmic
but slows down as the frozen steady state is approached.
Note that although the logarithmic behaviour is robust, the actual
values on the axes depend upon which of the various arbitrary choices
mentioned in the previous section have been made and hence have no physical
meaning.

\begin{figure}
\centerline{\psfig{file=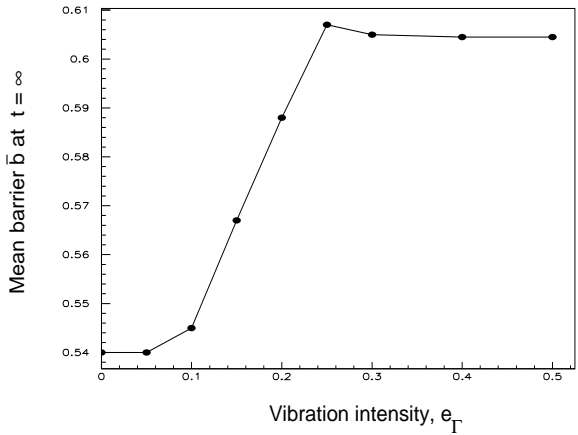,width=3in}}
\caption{The mean barrier height $\bar{b}$ in the final steady state
as a function of $e_{\Gamma}$. Note that since
\mbox{$\bar{b}\propto(\rho_{0}-\rho)^{-1}\sim\rho$} the
vertical axis can also be identified as the (approximate) density.
The simulations were performed on a
$10\times10\times10$ lattice and averaged over 1000 runs.
$b_{\rm init}=0.08$ and $b^{*}\approx0.25$.
}
\label{f:larget}
\end{figure}

Little has been said so far about initial conditions.
Before the first selection of the minimum
barrier $Q(b)$ is uniform
over the entire range $[0,1]$,
so that even a small $e_{\Gamma}$ will cause a significant amount
of reconfiguration.
This corresponds to a state of {\em minimum compactivity} which is very
difficult to attain experimentally.
For instance, there will always be a certain amount of background noise,
and the granules added later to the apparatus will impact upon those
already present, inevitably causing some compaction.
Instead, the experiments always started from a {\em slightly} compacted
state with a density fraction of $0.577\pm0.005$.
This initial compaction can be incorporated into the model by shifting
the time axis so that
the origin corresponds to when $G(t)$ first 
becomes greater than a parameter $b_{\rm init}>0$.
Values of $e_{\Gamma}\approx b_{\rm init}$ or less are too small
to cause any significant further compaction.
This is readily apparent in Fig.~\ref{f:larget}, where we have plotted
$\bar{b}$ in the limit~$t\rightarrow\infty$
against~$e_{\Gamma}$.
The line is flat for $e_{\Gamma}<b_{\rm init}$, increases linearly
for $b_{\rm init}<e_{\Gamma}<b^{*}$ and levels out again for higher 
$e_{\Gamma}$. This should be compared with the corresponding
experimental plot, which is Fig.~3 in~\cite{exp},
from which we estimate that $b^{*}$ corresponds to
$\Gamma\approx3$.

\begin{figure}
\centerline{\psfig{file=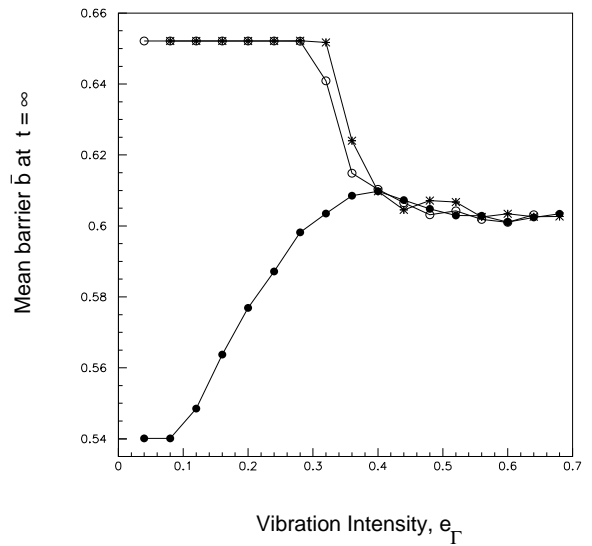,width=3in}}
\caption{Annealing curve for a $4\times4\times4$ lattice, for which
$b^{*}\approx0.38$. $e_{\Gamma}$ first increases from 0.04 to 0.68
in steps of 0.04 (filled circles), then decreases by the same
step size from 0.68 to 0.04
(open circles). Finally, $e_{\Gamma}$ is increased up to
0.68 again (asterixes). $b_{\rm init}$ was set at 0.08.
Each simulation was run until $t\approx156$, and the final plot was
averaged over 1000 such runs.}
\label{f:anneal}
\end{figure}

An apparently anomalous feature of Fig.~\ref{f:larget} is that the
highest densities are to be found, not for large $e_{\Gamma}$, as might be
expected, but instead for values of $e_{\Gamma}$ near the threshold
value~$b^{*}$.
This occurs because of finite size effects.
Recall that, for \mbox{$e_{\Gamma}>b^{*}$}, the barrier distribution
evolves to a state which is uniform for \mbox{$b>b^{*}$} with a tail for
\mbox{$b<b^{*}$}.
It is the very existence of this tail, which disappears in the
thermodynamic limit \mbox{$N\rightarrow\infty$},
that reduces the mean barrier $\bar{b}$ for finite systems.
When $e_{\Gamma}$ is slightly less than $b^{*}$ then, although the
uniform region is slightly broader, the
selection process can remove some of the barriers from the tail
permanently and so the net effect is to increase $\bar{b}$.
An even greater degree of compaction can be obtained if a system with
\mbox{$e_{\Gamma}>b^{*}$} is
first allowed to self-organise to the statistical steady state,
then $e_{\Gamma}$ is {\em slowly} reduced to zero to remove as much of
the tail as possible.
Quickly reducing $e_{\Gamma}$ will not give enough time for the selection
process to work before the system froze and so
$\bar{b}$ would barely change.
An example of this process is given in Fig.~\ref{f:anneal}, where to
accentuate the finite size effects a $4\times4\times4$ lattice was used.
Nowak~{\em et.al.} have produced similar plots from 
their experiments, which they
regard as a type of annealing process~\cite{exp2,exp3}.
They label the lower branch of the graph, when the intensity
of vibration is increased for the first time, as ``irreversible''.
In the language of our model, we prefer to call this the
{\em self-organising} branch.
The self-organising branch meets an upper {\em reversible} branch
around the point \mbox{$\Gamma^{*}\approx3$}.
This is to be expected since, as mentioned in the previous paragraph, this
value of $\Gamma$ corresponds to the threshold value $b^{*}$, that is,
the point at which the system can self-organise into the statistical
steady state.
According to the model, the change in density along the upper
branch is due to the effects of finite size, so there should be a
greater variation when larger beads are used in the same sized apparatus.
This is in agreement with the experiments
except for when the largest bead size was used~\cite{exp3}.
In this case, although the overall density variation was the greatest,
a disproportionately large amount of it occurred
along the self-organising branch,
possibly due to the cylinder walls aligning the beads into a highly
compact crystalline configuration.
Another feature observed in the experiments is that the threshold value
$\Gamma^{*}$ appears to increase when $\Gamma$ is updated more rapidly.
The model agrees with this and attributes it to the larger number of
steps that will take place before the system has had time
to self-organise.

For \mbox{$e_{\Gamma}>b^{*}$} the steady state is statistical in nature,
so another test for the model would be to compare the fluctuations
of $\bar{b}$ around its steady state value to the fluctuations in density
measured experimentally.
However, as previously mentioned, the argument relating $\bar{b}$ to
$\rho$ is not expected to hold for small changes.
The change in density caused by, say, a single reconfiguration event will
be sensitive to the exact positions of a large number of granules
at that instant in time.
The experimental plot of density fluctuations is
Gaussian in form~\cite{exp2},
indicative of the large number of independent factors involved.
A more revealing distribution is the power spectrum of density fluctuations,
$S(f)$, where the frequency  $f$ is measured in units of $(\rm{taps})^{-1}$.
Experimentally, $S(f)$ was found to obey the power law
\mbox{$S(f)\propto f^{-\delta}$}, with \mbox{$\delta=0.9\pm0.2$},
for a broad range of $f$.
Apart from finite size effects, the model predicts a power law
with $\delta=1$~\cite{BSRev}.
When large intensities of vibration were applied in the experiments,
the power law behaviour
was broken up by regions with $\delta=0$, $0.5$ or~$2$.
We cannot account for this and attribute it to the expected
breakdown of the model for large vibrations.

We end this section by briefly considering how the
model might also be applied to a set of related experiments.
Jaeger~{\em et.al.}~\cite{slope} have shown that the angle of
repose~$\theta(t)$ of granular media in a half-filled cylindrical drum
relaxes according to \mbox{$\theta(t)\sim\log t$} when vibrated.
Furthermore, they also demonstrated the existence of a threshold in
the intensity of vibration below which the relaxation was qualitatively
slower.
If we ignore the compaction process which presumably occurs
in the bulk of the pile, then the typical vertical separation between
granule centres is now proportional to $\tan\theta$, although
for the range of angles involved we can use
\mbox{$\tan\theta\approx\theta$} instead.
We can now repeat the argument given earlier for density and find
that the relationship between
the mean barrier and the slope is
\mbox{$\bar{b}\sim\theta$}, to first order.
Hence the model also predicts relaxation of
the form \mbox{$\theta(t)\sim\log t$} and the existence of the
threshold in the intensity of vibration.
However, we have reservations in applying the model to this new
geometry since it blatantly involves a global, albeit slow, movement of
granules over the surface, something which we have explicitly stated
the model does {\em not} cater for.
It should also be mentioned that other theoretical explanations for
this behaviour have already been given~\cite{mehta1,mehta2,slope}.

\section{Mean-Field Analysis}
\label{sec:mft}

The picture presented thus far can be extended by considering a
mean field version of the model which is open to quantitative analysis.
This simplified model exhibits many of the traits apparent in the
exact model, especially in the relaxation towards the
statistical steady state.
However, it behaves very differently in the steady state itself,
and we refer the reader elsewhere for analysis of the original model
in this much studied regime~\cite{BS,BSRev}.
The required mean field approximation is to be achieved in two stages.
First, all spatial definition is removed.
This means that, when the cluster with the smallest barrier in a
system of $N$ clusters is found and reconfigured, $K$ other clusters
are chosen at random from the remaining $N-1$
and their barriers given new values.
These $K$ clusters are equivalent to the adjacent clusters in the
original model, so for example $K=6$ corresponds to a 3 dimensional system.
The second simplification is to assume that $N$ is very large.
In this way the system can be described by continuous rather than
discrete variables, to within an error margin of $O(1/N)$.

For the first part of this section, the evolution of the system will
be described in terms of a time variable $\tau$ which
increases by $1/N$ between successive reconfigurations.
The inclusion of the variable time step given in~(\ref{timestep}) will be
postponed until later.
The system is described by the cumulative barrier
distribution~$C(b,\tau)$, which
is defined as the proportion of clusters with
barriers less than $b$ at time $\tau$
and is related to $Q(b,\tau)$ by

\begin{equation}
C(b,\tau)=\int_{0}^{b}Q(x,\tau)\,dx\:.
\end{equation}

\noindent{The time scale has been normalised to
one reconfiguration per cluster per unit~$\tau$,
so $C(b,\tau)$ evolves according to
}

\begin{equation}
\frac{\partial C(b,\tau)}{\partial\tau} =
-\theta(b-b_{\rm min}(\tau))-K\,C(b,\tau)+b\,(K+1)\:,
\label{rate}
\end{equation}

\noindent{where $b_{\rm min}(\tau)$ is the value of the minimum
barrier in the system at time~$\tau$ and
\mbox{$\theta(b)=1$} for \mbox{$b>0$} and 0 otherwise.
The removal of the minimum barrier has the effect of reducing
$C(b,\tau)$ for all values of~\mbox{$b>b_{\rm min}(\tau)$} but leaves
it unchanged for~\mbox{$b<b_{\rm min}(\tau)$}.
This is handled by the first term on the right hand side
of~(\ref{rate}).
In a similar manner, the second and third terms account for
the selection of the $K$ random nearest neighbours and the $K+1$
new barrier values, respectively.
It is straightforward to check that~(\ref{rate}) preserves
\mbox{$C(0,\tau)=0$}, \mbox{$C(1,\tau)=1$} and
\mbox{$C(b_{1},\tau)\geq C(b_{2},\tau)$} for~$b_{1}>b_{2}$,
for all values of~$\tau$.
}

The rate equation~(\ref{rate}) is not yet in a closed form because
it involves the unknown quantity~$b_{\rm min}(\tau)$.
We might naively try to write down a second equation giving
$b_{\rm min}(\tau)$ in terms of $C(b,\tau)$, perhaps something like
\mbox{$C(b_{\rm min}(\tau),\tau)=1/N$}.
However, it must be recalled that errors of $O(1/N)$ have already
been made in going from the discrete model to
this continuous description,
and so $C(b,\tau)$ cannot be used to this degree of accuracy.
Indeed, any attempt to define the minimum barrier within a continuum
framework is doomed to failure for this very reason.
We are forced to conclude that there can be no set of closed equations
in terms of $C(b,\tau)$.
All is not lost, however, since this problem can be partially
circumnavigated by use of the gap function $G(\tau)$.
As before, $G(\tau)$ is defined as the highest value that
$b_{\rm min}(\tau)$ has ever taken, or more formally,

\begin{equation}
G(\tau)=\sup_{0\leq z\leq\tau} b_{\rm min}(z)\:.
\end{equation}

\noindent{Values of $b$ greater than $G(\tau)$ must by definition
be greater than every value $b_{\rm min}$ has taken up to a time $\tau$.
This allows for~(\ref{rate}) to be simplified to
}

\begin{equation}
\frac{\partial C(b,\tau)}{\partial\tau} =
-(K\,C(b,\tau)+1)+b\,(K+1)\:,
\label{rate2}
\end{equation}

\noindent{for $b>G(\tau)$.
This can be solved by substituting
\mbox{$C(b,\tau)=\alpha(\tau)b+\beta(\tau)$} and comparing
coefficients of $b$.
With the initial condition \mbox{$C(b,0)=b$}
(so~$b_{\rm init}=0$), the result is
}

\begin{equation}
C(b,\tau)=b + \frac{b-1}{K}\left(1-e^{-K\tau}\right)
\label{cwitht}
\end{equation}

\noindent{The fact that $C(b,\tau)$ is linear means that the barrier
distribution $Q(b,\tau)$ is uniform for \mbox{$b>G(\tau)$},
as expected.
The solution~(\ref{cwitht}) holds from
\mbox{$b=1$} down to \mbox{$b\approx G(\tau)$},
where the continuum approximation starts
to break down and we have entered into
the asymptotic tail.
Since there are only $O(1/N)$ clusters in this tail,
the value of $G(\tau)$ will correspond to the point at
which $C(b,\tau)$ is zero,
ie.~\mbox{$C(G(\tau),\tau)=0$}.
Together with~(\ref{cwitht}) this allows for the time dependent form
of $G(\tau)$ to be found,
}

\begin{equation}
G(\tau)=\frac{ 1-e^{-K\tau} }{ K+1-e^{-K\tau} }\:.
\label{gap}
\end{equation}

\noindent{Ray and Jan have also found this result by an alternative
method~\cite{RayJan}.
The threshold value of $b$ in this mean field model is therefore
}

\begin{equation}
b^{*}=\lim_{\tau\rightarrow\infty}G(\tau)=\frac{1}{K+1} \:,
\end{equation}

\noindent{which is smaller than in the exact model.
}

In this approximation, the mean barrier height~$\bar{b}$ behaves
in the same way as the gap function.
This is because, to $O(1/N)$, there is no tail for
\mbox{$b<G(\tau)$} and the barrier distribution is
uniform for \mbox{$b>G(\tau)$}, so
\mbox{$\bar{b}(\tau)=(1+G(\tau))/2$}, which is just a
linear rescaling.
Hence we expect $G(\tau)$ to vary logarithmically with~$\tau$.
When the expression for $G(\tau)$ given in (\ref{gap}) is
plotted against $\log\tau$ it exhibits a linear
region similar to the exact model, but not
extending quite as close to the steady state.
The gradient of $G(\tau)$ in this log-linear plot is

\begin{equation}
\frac{{\rm d}G(\tau)}{{\rm d}(\ln\tau)} =
\tau\frac{{\rm d}G(\tau)}{{\rm d}\tau}=
\tau G'(\tau) \:.
\end{equation}

\noindent{The linear region occurs around the point where
the gradient is stationary, ie. when the second derivative is zero,
}

\begin{equation}
\frac{\rm d}{ {\rm d}(\ln\tau) }
\left( \frac{{\rm d}G(\tau)}{{\rm d}(\ln\tau)}\right ) =
\tau\left(G'(\tau)+\tau G''(\tau)\right)=0 \:.
\label{critpnt}
\end{equation}

\noindent{The solution with \mbox{$\tau=0$} corresponds to
the singularity in $\ln\tau$ and can be ignored.
Using~(\ref{gap}), the non-trivial solution is
}

\begin{equation}
\tau=\frac{1}{K}\tanh \frac{K}{2}(\tau+\tau_{0})\:,
\label{tanh}
\end{equation}

\noindent{where the constant \mbox{$\tau_{0}=(\ln(K+1))/K$}.
Since the slope is roughly constant in this region there is
no need to find the exact value of $\tau$ that
satisfies~(\ref{tanh}).
Instead we observe that, for large~$K$, the tanh function is
roughly equal to~1 for all \mbox{$\tau>0$},
so an approximate solution is \mbox{$\tau\approx1/K$} and hence the slope is
}

\begin{equation}
\left.
\frac{ {\rm d}G(\tau) }{ {\rm d}(\ln\tau)}
\right|_{\tau\approx\frac{1}{K}}
 \approx
\frac{Ke}{[(K+1)e-1]^{2}} \:.
\label{slope}
\end{equation}

We now turn to consider the effect of the variable timestep~$\delta t$
as defined in~(\ref{timestep}), which depends on $b_{\rm min}$
and~$e_{\Gamma}$.
The quantity $b_{\rm min}$ is unknown, but
we know from the discrete model that it fluctuates
between 0 and~$G(\tau)$ and therefore
substituting $G(\tau)$ for $b_{\rm min}(\tau)$
gives a qualitatively identical solution.
The new time scale is denoted by $t(\tau)$ and is defined by

\begin{equation}
\frac{ {\rm d}t }{ {\rm d}\tau } =
\exp\left\{\mu\left(
\frac{ e_{\Gamma} }{ e_{\Gamma}-G(\tau) }
\right)\right\}\:.
\label{ttau}
\end{equation}

\noindent{For small $\tau$, \mbox{$G(\tau)=\tau+O(\tau^{2})$}
and (\ref{ttau}) can be solved with the initial condition
\mbox{$t(0)=0$} to give
}

\begin{equation}
t(\tau)=e^{\mu}\left(
\tau+\frac{\mu}{2e_{\Gamma}}\tau^{2}+O(\tau^{3})
\right) \:,
\end{equation}

\noindent{which is linear up to
\mbox{$\tau=O(e_{\Gamma}^{\:\frac{1}{2}})$}.
The behaviour of $t(\tau)$ for large $\tau$ depends upon
whether $e_{\Gamma}$ is greater than, less than or equal
to the threshold value~\mbox{$b^{*}=\frac{1}{K+1}$}.
For \mbox{$e_{\Gamma}>b^{*}$},
\mbox{$G(\tau)\rightarrow\frac{1}{K+1}$}
as \mbox{$\tau\rightarrow\infty$} and consequently
}

\begin{equation}
t\sim\tau
\exp\left\{\mu\left(
\frac{ e_{\Gamma} }{ e_{\Gamma}-\frac{1}{K+1} }
\right)\right\} \:.
\end{equation}

\noindent{The time scale is stretched by a constant factor, but
otherwise the system approaches the same statistical steady state
as before.
For \mbox{$e_{\Gamma}<b^{*}$}, (\ref{ttau}) becomes singular
at the point \mbox{$\tau=\tau_{\rm crit}$}
at which \mbox{$G(\tau_{\rm crit})=e_{\Gamma}$}.
Since $\delta t$ diverges there are no more reconfigurations and the
system is in a frozen steady state.
The precise nature of this singularity can be found by substituting
\mbox{$\tau=\tau_{\rm crit}-\epsilon$} into~(\ref{ttau}),
with $\epsilon$ small and positive.
As \mbox{$\epsilon\rightarrow0$}, $t(\tau)$ diverges according to
}

\begin{equation}
\left.\frac{ {\rm d}t }{ {\rm d}\tau } \right|_{\epsilon\rightarrow0}
\sim
e^{A/\epsilon} \:,
\end{equation}

\noindent{where the constant}

\begin{equation}
A=\mu\frac{e_{\Gamma}}{ (1-e_{\Gamma})(1-(K+1)e_{\Gamma}) } \:.
\end{equation}

\noindent{Finally, for \mbox{$e_{\Gamma}=b^{*}$}
(\ref{ttau}) can be algebraically reduced to}

\begin{equation}
\left.\frac{ {\rm d}t }{ {\rm d}\tau }\right|_{\tau\rightarrow\infty}
\sim
\exp\left\{
\mu\frac{K+1}{K}e^{K\tau}
\right\} \:
\end{equation}

\noindent{for large $\tau$, which is divergent.}

Now that we have confirmed that $e_{\Gamma}$ has the same effect in the
mean field model as in the exact model, we need to see what it does
to the rate of logarithmic decay.
This is straightforward for~\mbox{$e_{\Gamma}\gg b^{*}$} since

\begin{equation}
t=e^{\mu}\tau+O(e_{\Gamma}^{\:-1}) \:,
\end{equation}

\noindent{so to first order in~$e_{\Gamma}^{\:-1}$
the time scale is just stretched by a constant factor,
which does not alter the gradient in a log-linear plot.
This means that slope of $G(t)$~vs~$\log t$
is the same as the slope of $G(\tau)$~vs~$\log\tau$
and~(\ref{slope}) can be used without modification.
For instance, in the exact system with large~$e_{\Gamma}$
the slope is approximately 0.048 in 3 dimensions,
whereas the value predicted by~(\ref{slope}) for
\mbox{$K=6$} is 0.050.
}

Modifying (\ref{slope}) to incorporate
\mbox{$e_{\Gamma}<\infty$} is troublesome and
we have been unable to derive a general formula.
Nonetheless there is still some hint of a correspondence
between this analysis and the experiments.
In~\cite{exp} Knight~{\em et.al.} introduce a parameter~$\tau$
which we call $\tau_{\rm exp}$ so as not to confuse it with
our~$\tau$.
$\tau_{\rm exp}$~gives a rough
measure of the time scale of the relaxation process.
We tentatively equate this to the quantity
\mbox{$ {\rm d}t/{\rm d}\tau $}, and indeed the
experimental plot of $\tau_{\rm exp}$~vs~$\Gamma$
looks similar to the form
of \mbox{$ {\rm d}t/{\rm d}\tau $}
given in~(\ref{ttau}).
However, this
is not a robust feature of the model and so it is
impossible to come to any concrete conclusions.
The experimental data also shows a noticeable change in behaviour
for small~$\Gamma$.
This could be caused the system entering into the frozen steady
state before the logarithmic relaxation has had a chance to take hold,
ie. when~\mbox{$\tau_{\rm crit}\ll\frac{1}{K}$}, although it could
just be the effect of the initial compaction.

Finally, we demonstrate how this analysis can
be extended to incorporate
energy dissipated by a reconfiguring cluster
to its nearest neighbours.
Suppose that each adjacent cluster receives an energy $e_{\rm diss}$
and immediately reconfigures if its barrier is
smaller than this, dissipating a further energy $e_{\rm diss}$
to each of its neighbours, and so on.
Using the same mean field approximations as before,
the net effect of this avalanche between perturbations
is to increase the
number of barriers that change value at each time step.
Of the $K$ random nearest neighbours,
$Ke_{\rm diss}$ will immediately reconfigure and so the
total number of new barriers per time step~${\rm d}\tau$ is now

\[
K+K(Ke_{\rm diss})+K(Ke_{\rm diss})^{2}
+K(Ke_{\rm diss})^{3}+\ldots
\]

\begin{equation}
= \frac{K}{1-Ke_{\rm diss}} \:,
\end{equation}

\noindent{for $e_{\rm diss}<\frac{1}{K}$. Larger values of
$e_{\rm diss}$ are unphysical since they result in perpetual
reconfiguration.
The new rate equation for $C(b,\tau)$ is
}

\[
\frac{\partial C(b,\tau)}{\partial\tau} =
-\theta(b-b_{\rm min}(\tau))
-\frac{K}{1-Ke_{\rm diss}}
\]

\begin{equation}
+\left(1+\frac{K}{1-Ke_{\rm diss}}\right)b \:,
\end{equation}

\noindent{which can be solved as before to give}

\begin{equation}
C(b,\tau)=b+\frac{b-1}{K}
(1-Ke_{\rm diss})
\left(1-\exp\left[-\frac{K\tau}{1-Ke_{\rm diss}}\right]\right)
\end{equation}

\noindent{for $b>G(\tau)$.
This is the same as the solution already given in~(\ref{cwitht})
except that $K$ has been
replaced by the effective number of random nearest neighbours
\mbox{$K/(1-Ke_{\rm diss})$}.
The time scale is similarly stretched by the constant factor
\mbox{$1-Ke_{\rm diss}$}.
Hence the inclusion of energy dissipation in this manner
does not alter the behaviour of the system, nor does it change
the slope of $G(\tau)$ in a log-linear graph.
}

\section{Summary and discussion}
\label{sec:summ}

We have presented a theoretical model for the compaction
of granular materials by low intensity perturbations
which appears to agree well with a range of experimental results.
This includes the logarithmic relaxation, the effect of varying the
intensity of vibration resulting
in a so-called ``annealing'' curve, and
the power spectrum of density fluctuations in the steady state.
We have segmented the granular media
into local subsystems or clusters
which represent ensembles of granules
that collectively reconfigure.
Associated with each cluster is a potential energy
barrier against reconfiguration.
Whenever a perturbation gives a cluster enough energy
to cross over its barrier into a new configuration,
nearby clusters are disrupted and
their barriers take on new values.
The system behaviour is dominated by this dynamical interaction
between clusters and fine detail such as the choice of
distribution for the barrier values makes little or no difference.
Indeed, it is this very robustness that leads us to hope that
the model might correctly describe the mechanism underlying
the compaction process, despite its algorithmic simplicity.

It has been suggested that standard statistical mechanics can be
applied to granular materials if the fundamental quantities
involved are suitably reinterpreted~\cite{statmech1,statmech2}.
Volume plays the role of energy,
and the quantity conjugate to volume is known as {\em compactivity},
which is the analogue of temperature.
The compactivity is infinite when the system is
at its maximum volume and zero when it is at its minimum.
Our model can also be described in terms of
volume rather than energy since the external perturbations
increase the volume of the system as well as its energy.
Hence we can assign a volume barrier to each cluster which must
be exceeded for reconfiguration to take place.
In this way, we can see the beginnings of a link to the modified
statistical mechanics, perhaps with the barriers being
in some way related to the compactivity.
This is just speculation, however, and further investigation is required.
There are also be many ways in which the model can be enhanced
make it more physically realistic.
For instance, the model is currently isotropic, but real granular media
exhibits a density gradient with the densest regions near the bottom.

There is another way to compact granules into a smaller
volume, and that is simply to apply a uniform pressure.
This forces the granules to rearrange into a higher density state,
as with the perturbation-induced compaction studied in this paper,
although the granules are now also subject to deformation and fracturing.
A theoretical model for compaction by applied pressure has been proposed
which treats the media as being comprised of a number of subsystems,
each of which is associated with a pressure barrier~\cite{pressure}.
This obviously bears some similarity to the approach we have
adopted in constructing our model.
A crucial difference is that the subsystems in
the pressure model do not interact
and the values for the barriers are simply drawn from a
suitable distribution.
In our model, the choice of distribution is unimportant and it
is the dynamical interactions between subsystems that dominates
the system behaviour.
It would be interesting to see if the interacting cluster picture
can be applied to this or any other experimental situation
involving granular materials.

\section*{acknowledgment}

We would like to thank Prof.~Heinrich~Jaeger for useful discussions
concerning the experiments and for supplying us with
preprints~\cite{exp2,exp3}.

\end{multicols}


\begin{references}

\bibitem [*] {em1} Electronic address: David.Head@brunel.ac.uk

\bibitem [\dag] {em2} Electronic address: G.J.Rodgers@brunel.ac.uk

\bibitem{rev1} H.~M.~Jaeger and S.~R.~Nagel, Science~{\bf 255}, 1523 (1992).

\bibitem{rev2} H.~M.~Jaeger, S.~R.~Nagel and R.~P.~Behringer,
Rev.~Mod.~Phys.~{\bf~68}, 1259 (1996).

\bibitem{industrial} P.~E.~Evans and R.~S.~Millman in {\em Perspectives in 
Powder Metallurgy, Vol 2 : Vibratory Compacting} (Plenum Press,
New York, 1967).

\bibitem{mehta1} Anita~Mehta, Physica~A~{\bf 186}, 121 (1992).

\bibitem{mehta2} Anita~Mehta in {\em Granular Matter: An Interdisciplinary
Approach}, edited by Anita~Mehta (Springer-Verlag, New York, 1994).

\bibitem{mehta3} G.~C.~Barker and Anita~Mehta, Phys.~Rev.~E~{\bf 47},
184 (1993).

\bibitem{exp} J.~B.~Knight, C.~G.~Fandrich, C.~N.~Lau, H.~M.~Jaeger and 
S.~R.~Nagel, Phys.~Rev.~E~{\bf 51}, 3957 (1995).

\bibitem{linz} S.~J.~Linz, Phys.~Rev.~E {\bf 54}, 2925 (1996).

\bibitem{rsa1} E.~Ben-Naim, J.~B.~Knight and E.~R.~Nowak,
{\em ``Slow Relaxation in granular compaction''}, preprint,
{\em cond-mat/9603150}.

\bibitem{rsa2} P.~L.~Krapivsky and E.~Ben-Naim, J.~Chem.~Phys.~{\bf 100},
6778 (1994).

\bibitem{exp2} E.~R.~Nowak, J.~B.~Knight, E.~Ben-Naim, H.~M.~Jaeger and
S.R.~Nagel, {\em ``Density fluctuations in vibrated granular
materials''}, to be published in Phys.~Rev.~E.

\bibitem{de_Gennes} T.~Boutreux and P.~G.~de~Gennes, 
{\em ``Compaction of granular materials: a free volume model''}, preprint.

\bibitem{tetris} E.~Caglioti, V.~Loreto, H.~J.~Herrmann and M.~Nicodemi,
{\em ``A tetris-like model for the compaction of dry granular media''},
preprint, {\em cond-mat/9705195}. Submitted to Phys.~Rev.~Lett.

\bibitem{frus1} M.~Nicodemi, A.~Coniglio and H.~J.~Herrmann, Phys.~Rev.~E
{\bf 55}, 3962 (1997).

\bibitem{BS} P.~Bak and K.~Sneppen, Phys.~Rev.~Lett.~{\bf 71}, 4083 (1993).

\bibitem{BSRev} M.~Paczuski, S.~Maslov and P.~Bak,
Phys.~Rev.~E~{\bf 53}, 414 (1996).

\bibitem{beadheap} P.~Evesque and J.~Rajchenbach,
Phys.~Rev.~Lett.~{\bf 62}, 44 (1989).

\bibitem{exp3} E.~R.~Nowak, J.~B.~Knight, M.~Povinelli, H.~M.~Jaeger
and S.~R.~Nagel, {\em ``Reversibility and irreversibility in the packing
of vibrated granular material''}, to be published in Powder Technol.

\bibitem{slope} H.~M.~Jaeger, C.~Liu and S.~R.~Nagel,
Phys.~Rev.~Lett.~{\bf 62}, 40 (1989).

\bibitem{RayJan} T.~S.~Ray and N.~Jan, Phys.~Rev.~Lett~{\bf 72}, 4045 (1994).

\bibitem{statmech1} Anita~Mehta and S.~F.~Edwards, Physica~A~{\bf 157},
1091 (1989).

\bibitem{statmech2}  S.~F.~Edwards in {\em Granular Matter: An
Interdisciplinary Approach}, edited by Anita~Mehta
(Springer-Verlag, New York, 1994).

\bibitem{pressure} V.~M.~Kenkre, M.~R.~Endicott, S.~J.~Glass and A.~J.~Hurd,
J.~Am.~Ceram.~Soc.~{\bf 79}, 3045 (1996).

\end{references}
\end{document}